\PassOptionsToPackage{table}{xcolor}
\documentclass[format=acmsmall, review=false, screen=true]{acmart}

\usepackage{booktabs} % For formal tables
\usepackage[ruled]{algorithm2e} % For algorithms
\usepackage{natbib}
\usepackage{enumitem}
\usepackage{tabularx}
\usepackage{siunitx}
\usepackage{multirow}
\usepackage{subcaption}
\usepackage[justification=centering]{caption}
\newcolumntype{Y}{>{\centering\arraybackslash}X}

\SetAlFnt{\small}
\SetAlCapFnt{\small}
\SetAlCapNameFnt{\small}
\SetAlCapHSkip{0pt}
\IncMargin{-\parindent}

\acmJournal{PACMHCI}
\acmVolume{2}
\acmNumber{CSCW}
\acmArticle{80}
\acmYear{2018}
\acmMonth{11}
\acmPrice{15.00}
\acmDOI{10.1145/3274349}
\setcopyright{acmlicensed}
\received{April 2018}
\received[revised]{July 2018}
\received[accepted]{September 2018}

\begin{document}
% Title portion. Note the short title for running heads
\title[``The Perfect One'']{``The Perfect One'': Understanding Communication Practices and Challenges with Animated GIFs}

\author{Jialun "Aaron" Jiang}
\affiliation{
  \institution{University of Colorado Boulder}
  \department{Department of Information Science}
  \streetaddress{ENVD 201, 1060 18th St.}
  \city{Boulder}
  \state{CO}
  \postcode{80309}
  \country{USA}}
\email{aaron.jiang@colorado.edu}

\author{Casey Fiesler}
\affiliation{
  \institution{University of Colorado Boulder}
  \department{Department of Information Science}
  \streetaddress{ENVD 201, 1060 18th St.}
  \city{Boulder}
  \state{CO}
  \postcode{80309}
  \country{USA}}
\email{casey.fiesler@colorado.edu}

\author{Jed R. Brubaker}
\affiliation{
  \institution{University of Colorado Boulder}
  \department{Department of Information Science}
  \streetaddress{ENVD 201, 1060 18th St.}
  \city{Boulder}
  \state{CO}
  \postcode{80309}
  \country{USA}}
\email{jed.brubaker@colorado.edu}

\renewcommand{\shortauthors}{J. A. Jiang et al.}

\begin{abstract}
Animated GIFs are increasingly popular in text-based communication. Finding the perfect GIF can make conversations funny, interesting, and engaging, but GIFs also introduce potentials for miscommunication. Through 24 in-depth qualitative interviews, this empirical, exploratory study examines the nuances of communication practices with animated GIFs to better understand why and how GIFs can send unintentional messages. We find participants leverage contexts like source material and interpersonal relationship to find the perfect GIFs for different communication scenarios, while these contexts are also the primary reason for miscommunication and some technical usability issues in GIFs. This paper concludes with a discussion of the important role that different types of context play in the use and interpretations of GIFs, and argues that nonverbal communication tools should account for complex contexts and common ground that communication media rely on.
\end{abstract}





\maketitle
%\renewcommand{\shortauthors}{G. Zhou et al.}

\section{Introduction}

The SMS notification on Jane's phone chimes almost constantly. Throughout the day she keeps in touch with dozens of friends, sharing messages, images, and more recently, animated GIFs. Jane just got a job offer from a local company and wants to share how happy she is. She opens her messaging app, finds one of her friends, types ``me right now.'' She then brings up the built-in GIF keyboard, searches ``happy,'' and attaches a GIF with two cartoon characters dancing and laughing. She hits the ``Send'' button, and her friend's confused reply arrives almost immediately: ``Brian and Stewie? 4:20 much? I didn't know you smoked pot.'' Jane stares at her phone in horror, as she had just sent the same GIF to her mother.

Animated GIFs have long been an important part of Internet culture and visual design. Created in 1987 as an image file format with additional support for animation sequences, GIFs grew to become an integral part of the visual Internet during the nineties, flashing ``under construction'' signs and spinning email icons on personal webpage \cite{nooney_brief_2014}. After declining in use for a better part of a decade and seen as a relic of outdated web design, GIFs have resurged through their reintroduction on MySpace and subsequent adoption by online communities like Reddit and Tumblr \cite{miltner_never_2017}. 

Although the file format has remained the same, GIFs are now used in a different way, focused on the ability to put snippets of videos into a single file.\footnote{In this paper, we focus on this currently popular style of GIF, and we refer to animated GIFs as ``GIFs'' for the sake of brevity.}  GIFs often contain multiple layers of meaning and can perform a wide range of emotions, and have gained significant cultural prominence \cite{ash_sensation_2015,miltner_never_2017}. These silent, short, usually low-resolution clips now pervade online communication, and have become part of the standard functionality of many social media sites and instant messaging applications. In fact, prior research found GIFs were the most engaging kind of media on Tumblr in terms of likes and reblogs \cite{bakhshi_fast_2016}. 

As noted by \emph{The New York Times}, GIFs are ``a way to relay complex feelings and thoughts in ways beyond words and even photographs'' \cite{isaac_for_2015}. However, the abundance of meanings in GIFs can be a double-edged sword. Just as we might misinterpret the meaning behind a facial expression, or an emoji \cite{miller_blissfully_2016}, interpretation poses a problem with GIF communication as well. Prior work has revealed that, similar to emojis, people can have diverse interpretations of the same GIF \cite{jiang_understanding_2017}. As misinterpretations can cause problems in communication, it is critical that we understand the reasons for misinterpretation, the ways in which miscommunication happens because of it, and the factors at play in people's communication with GIFs in general. Therefore, in this study, we investigate people's experiences of miscommunication when using animated GIFs, and by situating these experiences within their broader communication practices, we also explore the nuanced ways people engage with a novel communication media.

Across 24 in-depth qualitative interviews with GIF users, we heard many stories of miscommunication like Jane's above. Through an analysis of these interviews, we identify ways that people use GIFs, the impact of technical functionality on GIF use, and the nuanced relationship between context in GIFs and communication---while the context communicated through GIFs can enhance communication, it also introduces barriers to use and paths to miscommunication. People want to find the perfect GIFs to convey their emotions to their communication partners, but they often fail to do so due to the lack of consideration of context in the design of the platform.  Through a discussion of how GIF use and interpretations are shaped by the interpersonal and communal contexts that the media affords, we contribute a deeper understanding of communication practices with a novel communication media, and discuss design opportunities that prioritize context as a key consideration.

\section{Related Work}
To provide a foundation for understanding communication with GIFs, we summarize literature on nonverbal computer-mediated communication (CMC) and common ground, and then discuss prior work on communication and miscommunication with emoticons and emoji as examples of widely used nonverbal media. We conclude with a discussion of existing research on GIFs.

\subsection{Nonverbal Communication in CMC}
GIFs, like many forms of non-verbal communication, have the potential to introduce ambiguity and miscommunication. In early CMC research, Kiesler et al. and Sproull et al. \cite{kiesler_social_1984,sproull_reducing_1986} noted that CMC conveyed fewer contextual and nonverbal cues, resulting in a scarcity of social context information. However, later research has acknowledged the availability of nonverbal cues and their ability to significantly affect communicators' perceptions in CMC. In the development of social identity-deindividuation (SIDE) theory, Lea and Spears argued that nonverbal cues were an important source of CMC information used to form impressions when communicating \cite{lea_paralanguage_1992}. In the absence of interpersonal cues, communicators form impressions from whatever limited cues are available.  

Other CMC theorists have argued that the absence of contextual cues in CMC can be overcome. Social Information Processing Theory argues that over time, and after sufficient exchanges, communicators develop sufficient personal and relational information to negate the scarcity of cues \cite{walther_interpersonal_1992}. Building on this theory, the hyperpersonal model argues that CMC can surpass face-to-face communication as it allows communicators to strategically manage impressions. Message receivers have idealized perceptions of senders, not only due to the receivers' over-reliance on minimal cues such as word choice and punctuation usage, but also because the senders can selectively present themselves \cite{walther_computer-mediated_1996}. This loop of perception intensification enables CMC to exceed the limits of interpersonal communication---it became ``hyperpersonal.''

\subsection{Common Ground}
While theories such as SIDE and the hyperpersonal model shed light on how people leverage nonverbal cues to form impressions, Clark's theory of common ground \cite{resnick_grounding_1991} provides insights into how these cues are interpreted.  Clark described the interpersonal contexts and the media contexts that GIFs heavily reference as the \emph{personal} and \emph{communal} common grounds, as well as how communicators establish mutual knowledge around these types of common grounds \cite{clark_definite_1981}. 

While Clark considers many steps in the process of establishing a shared understanding---from noticing a speech act to understanding its meaning---our work largely focuses on how people make sense of and interpret GIFs' meanings. A successful contribution to the conversation consists of two phases: \emph{presentation}, wherein the speaker presents an utterance, and \emph{acceptance}, wherein the addressee provides evidence that the utterance is understood. According to Clark, presentation can be highly complex; this is echoed in our findings, wherein communicators must overcome much technical overhead to present the GIF as intended, and sometimes the overhead does not allow communicators to do so at all. Even when the GIFs are successfully presented, the multiple kinds of contexts that GIFs reference introduce potentials for problems in acceptance---in our case, miscommunication with GIFs. Most miscommunication happens when people receive the message but do not understand its meaning, but in some cases people do not see the message as a form of communication at all (referred to by Clark as State 2 and State 1, respectively).

People try to minimize their collaborative effort to establish common ground, but the effort required and the availability of grounding techniques dramatically differs between different media \cite{resnick_grounding_1991}. As we will discuss, GIFs as utterances can lead to a complicated presentation phase because of their multiple layers of meanings and references. On the one hand, participants viewed this complicated presentation as a feature of GIFs, as it reflects their various kinds of common ground. On the other hand, participants also described miscommunication when they found their communication partners failed to provide evidence that they completely understood the presentation from GIFs.

\subsection{Emoticon and Emoji}
Given the limited scholarship on animated GIFs, we leverage scholarship focused on emoticons and emoji, especially as they are also nonverbal media commonly used to supplement text communication.

In today's CMC, some of the most common nonverbal cues are shared through emoticons and emoji. Prior work has addressed communication practices and experiences with both. Emoticons are typographic symbols that appear sideways as resembling facial expressions \cite{walther_impacts_2001}. People use emoticons to speed up communication and express emotions more easily than using pure text \cite{huang_exploring_2008}. Additionally, emoticons are also used for tone management \cite{dresner_functions_2010} and punctuation \cite{lanchantin_case_2012}. Emoticons tend to reinforce the interpretations of messages but their impact on interpretations is generally overwhelmed by accompanied verbal messages \cite{walther_impacts_2001}. While interpretations of emoticons seem consistent within individual cultures, different cultural contexts can impact both the use and the interpretations of emoticons \cite{park_emoticon_2013,park_cross-cultural_2014}.

In the recent years, emoji have become one of the most popular form of nonverbal media in online text communication \cite{noauthor_emojineering_2015}. Emoji are pictographs that depict various genres, such as facial expressions, common objects, etc. \cite{noauthor_emoji_2017}. People use emoji in diverse ways: to provide additional emotional and situational information; change conversational tones; hide their true feelings; and maintain conversations and relationships with communication partners \cite{cramer_sender-intended_2016,kelly_characterising_2015}, which are also similar to the ways people use stickers \cite{lee_smiley_2016}.  Across cultures, people use emoji in similar ways \cite{zhou_goodbye_2017},  but the specific emoji they use differ \cite{lu_learning_2016}. Compared to emoticons, people also use emoji in inventive ways that differ from their original meanings. For example, \cite{kelly_characterising_2015} noted that emoji could take on special meanings within particular relationships, such as using the eggplant, peach, and taco emoji for sexual connotations \cite{highfield_instagrammatics_2016}, and \cite{wijeratne_word_2016} found that street gang members used the fuel pump emoji in the context of selling or consuming marijuana. 

However, these inventive ways to use emoji also mean that interpreting them tends to be inconsistent. People have varied interpretations of emoji both within-platform and cross-platform \cite{miller_blissfully_2016,tigwell_oh_2016}, and in both cases, accompanying text did not disambiguate emoji \cite{miller_understanding_2017}. Following this research, Miller et al. speculated that a reason that text did not disambiguate was the limited amount of common ground in their study \cite{miller_understanding_2017}.

In sum, while misinterpretation with emoticons seemed rare, the more recent phenomenon of emoji introduced a good degree more misinterpretation and potential for miscommunication. While we do not know of any research that systematically looks at this discrepancy or the reason for it, we speculate the higher potential for miscommunication can be attributed to the fact that they proliferate the repertoire of potential emotions, and they depict a greater variety of objects and activities \cite{noauthor_emoji_nodate}, in addition to people's more inventive ways to use them.  GIFs, as a new nonverbal media commonly used to supplement text communication, surpass emoji on all these dimensions: GIFs have more details and offer a greater variety of depictions---they are essentially short video clips. The same GIF can also be inventively used in various contexts. Therefore, we suspect GIFs also suffer the problem of misinterpretation and miscommunication, and this research seeks to investigate and unpack the nuance of miscommunication with GIFs.

\subsection{Animated GIFs}
Prior work on GIFs has focused on their linguistic functions. People use GIFs as co-speech demonstrations that depict emotions or actions, or as affective responses to prior messages produced by their communication partners \cite{tolins_gifs_2016}. GIFs also express emotions \cite{bakhshi_fast_2016,bourlai_multimodal_2014,miltner_never_2017}, and use movement, color, and repetition to create sensations and affect \cite{ash_sensation_2015}. Like emoji, GIFs are also polysemic, offering different meanings and interpretations to different audiences, and their meanings are often created within the context of a community \cite{miltner_never_2017}. GIFs also uniquely often derive from excerpts of movies or TV shows, reflecting cultural knowledge of this source material \cite{miltner_never_2017,nooney_brief_2014}. These characteristics make GIFs significantly more likely to be liked or reblogged than text, photos, and videos on Tumblr, according to Bakhshi et al. \cite{bakhshi_fast_2016} 

However, people also interpret GIFs in inconsistent ways. Previous work found that GIFs, as a form of GPOY, or ``gratuitous picture of yourself'' posts, were used in interpretively flexible ways \cite{tiidenberg_sick_2017}. Through a survey, Jiang et al. also showed people had diverse interpretations of GIFs, and that the duration of the animation as well as the overall emotion of the GIFs contributed to the variation \cite{jiang_understanding_2017}. 

While GIFs' capacity for multiple meanings and cultural references make them an appealing media to use, these same characteristics also introduce potential for miscommunication. In this work, we take a close look at how GIFs both positively and negatively impacted grounding in participants' communication, and focus on participants' nuanced practices surrounding GIFs.

\section{Methods}
In this exploratory qualitative study, we conducted semi-structured, in-depth interviews \cite{seidman_interviewing_2006} with 24 participants to understand their communication experiences with GIFs. After receiving approval from our Institutional Review Board (IRB), we proceeded with recruiting participants from public postings on social media, as well as online communities such as Reddit and Tumblr. We encouraged participants to share the recruitment posts, resulting in a snowball sample. There were no participation restrictions beyond a requirement that participants have any experience in using GIFs, be at least 18 years of age, and be able to conduct the interview in English.   

We conducted interviews remotely via voice communication tools (e.g., phone, Skype, or Google Hangouts; $n=19$) or in person ($n=5$), according to participants' preferences. Interviews ranged in length from 30 to 90 minutes, and all interviews were audio recorded. All participants lived in the United States at the time of interviews, with ages ranging from 18 to 36 ($M=26.6$, $SD=6.1$). Nine participants identified as male, and fifteen as female. Detailed information about the participants is shown in Table \ref{tab:demographics}.

\begin{table}
\begin{center}
\caption{Participant Demographics and Platforms of GIF Use}
\label{tab:demographics}
\begin{tabular}{cccc}
Participant ID & Age & Gender & Platform of GIF Use \\
\midrule
P01            & 20  & M      & iMessage, Facebook Timeline, Reddit                           \\
P02            & 27  & M      & Facebook Messenger, Twitter                                   \\
P03            & 21  & M      & Android Messages, Textra, Reddit, Tinder                      \\
P04            & 18  & F      & iMessage, Tumblr                                              \\
P05            & 19  & M      & iMessage, Twitter                                             \\
P06            & 22  & F      & Wechat                                                        \\
P07            & 26  & F      & Email, Facebook timeline, Tumblr, Reddit                      \\
P08            & 32  & M      & Google Hangouts, Twitter, Facebook timeline, Tumblr           \\
P09            & 18  & F      & Facebook Messenger, iMessage, Tumblr                          \\
P10            & 35  & F      & Google Hangouts, Facebook Messenger, Relay(discontinued)      \\
P11            & 21  & F      & Android Messages, Tumblr                                      \\
P12            & 35  & F      & iMessage, Whatsapp, Twitter                                   \\
P13            & 33  & M      & Twitter, Google Hangouts, Facebook timeline, Tumblr, iMessage \\
P14            & 35  & F      & Facebook timeline, Twitter, online forum                      \\
P15            & 32  & F      & Twitter, iMessage                                             \\
P16            & 29  & F      & iMessage, Email                                               \\
P17            & 21  & F      & Google Hangouts, Facebook Messenger, Tumblr, Email            \\
P18            & 30  & M      & Android Messages, Email, Tumblr, Reddit                       \\
P19            & 28  & F      & Google Hangouts, Twitter                                      \\
P20            & 27  & F      & iMessage, Facebook timeline, Twitter                          \\
P21            & 30  & M      & Email, iMessage, Google Hangouts, Facebook Messenger          \\
P22            & 22  & M      & iMessage, Facebook Messenger, Reddit                          \\
P23            & 29  & F      & Email, iMessage                                               \\
P24            & 36  & F      & Twitter, Facebook Messenger   \\   
\bottomrule

\end{tabular}
\end{center}
\end{table}

Our interviews were comprised of three portions. We started by asking participants about their use of GIFs, platforms on which they used GIFs, and occasions in which they would or would not use GIFs. Our participants used GIFs across multiple platforms, including email, instant messaging, and social media and online communities. Most participants (21 out of 24) in our study had extensive experience with GIFs and used them daily; the other 3 participants used GIFs multiple times a week.

In the open-ended portion of our interview, we asked participants to share scenarios from their own lives, specifically focusing on communication and miscommunication experiences when using GIFs. For each example that participants shared with us, we asked for situational information, such as the surrounding conversation, platform, and rationale behind use. 

\begin{figure}
  \centering
  
  \begin{minipage}[b]{.45\textwidth}
  	\centering
    \includegraphics[width=.9\textwidth]{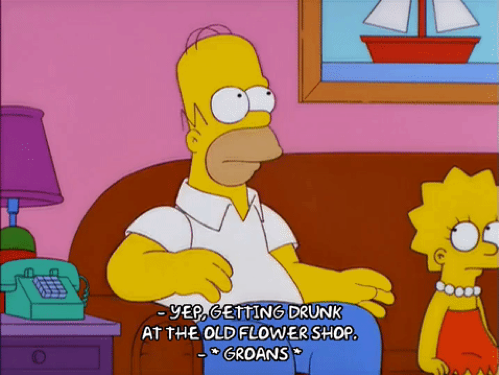}
    \caption{A snapshot of an example GIF from \emph{The Simpsons} we showed participants. \\ Retrieved from http://gph.is/1pqEncu}
    \label{fig:1}
  \end{minipage}\hfill
  \begin{minipage}[b]{.45\textwidth}
  	\centering
    \includegraphics[width=.9\textwidth]{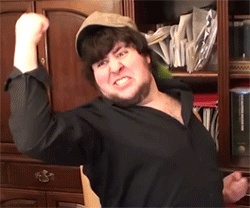}
    \caption{A snapshot of an example GIF of JonTron we showed participants. \\ Retrieved from http://gph.is/18AQbOM}
    \label{fig:2}
  \end{minipage}
\end{figure}

At the end of the interviews, we asked participants to give their interpretations of specific GIFs (e.g., Fig. \ref{fig:1}) for which prior work has found interpretations vary \cite{jiang_understanding_2017}. To ground their interpretations, we also asked participants to share potential use cases for these GIFs.

We conducted thematic analysis on the interviews \cite{braun_using_2006}, leveraging coding techniques described in \cite{saldana_coding_2009}. We began analysis by transcribing and anonymizing all interviews. Next, each author individually engaged in inductive open coding of a subset of the interview transcripts. During weekly team meetings, we iteratively discussed and revised these codes until we reached consensus. We clustered similar codes and identified emergent key themes.

\section{Findings}

Our findings from the data fall into three major themes: appeal, technical functionality, and context. In this section, we first discuss the reasons that GIFs appealed to participants. Then, we talk about different technical functionalities that facilitated or hindered participants' GIF use. Lastly, we discuss in detail how different types of context impact the ways people use and interpret GIFs. 

\subsection{The Appeal of Animated GIFs}
Participants described various ways to use GIFs. Throughout our interviews, we found that participants used GIFs as a way to enhance their interpersonal communication.
The most commonly cited motivation for using GIFs (18 out of 24 participants) was to convey emotion:

\begin{quote}
\textit{I...would plug in GIFs to like, convey specific emotion or a mood, and post reaction GIFs to other people's post, to kind of convey what they thought. (P04)}
\end{quote}

Participants used GIFs instead of text when they were unable to adequately express their emotions with text, and we found that GIFs provided participants with a wider range of emotions than text: 

\begin{quote}
\textit{[I use GIFs] in text conversations and to interpret something that I can't type, if that makes sense. If I can't type it, word for word, I'll find a GIF for it. (P05)}
\end{quote}

P05's quote captures how participants throughout our data turned to GIFs to express nuanced ideas and emotions that would be difficult to express via text alone. GIFs, however, provide more than just a wider range of emotions. Participants actively contrasted GIFs with emoji, and felt that GIFs better represented the affect they wanted to share:

\begin{quote}
\textit{I never felt emojis were worth it, like it was just a few extra letters to say ``I love you'' instead of writing a heart, but I feel like the emotion that is carried with a GIF can express more in terms of human understanding than just words...I feel like something that GIFs can express that you just can't express quickly in words, which I feel like you can with emoji. (P21)}
\end{quote}

While emoji can function simply as replacements for text, according to P21, participants did not think the same for GIFs. Even when they could use words to describe their intentions, participants chose GIFs because they could make messages humorous or eye-catching, even in simple ways:

\begin{quote}
\textit{It's like, if I just want to say like ``okay,'' ``yeah,'' or ``no'' like in a funny way, I'll just look it up, and have a GIF like somebody shaking their head, or doing something stupid. (P03)}
\end{quote}

\begin{quote}
\textit{...when I'm sending emails to like my colleagues or my friends... [I use GIFs] to make it more interesting, especially if I'm trying to grab someone's attention. [The] moving image definitely gets people to, I think, pay attention to the email a little bit more. (P06)}
\end{quote}

Participants also described using GIFs as communication tactics to initiate and maintain conversations, similar to those used in face-to-face communication. Participants used GIFs as ``a conversation starter,'' or pre-sequence \cite{sacks_lectures_1995}, similar to saying ``How's it going?'' to start a conversation; GIFs can also serve as continuers \cite{goodwin_between_1986} to fill in the blanks and maintain conversations:

\begin{quote}
\textit{Sometimes I'll just look up completely random things and send them to people like randomly as like a conversation starter. (P03)}
\end{quote}

\begin{quote}
\textit{But sometimes you want to just send something that says, you know, ``I read what you wrote.'' Like I don't really have anything to say about that, but whatever. You know, I'm still talking to you, so sometimes I would send a GIF for that purpose. (P22)}
\end{quote}

While our participant noted that emoji were also useful for phatic communication, confirming prior work \cite{kelly_characterising_2015}, participants turned to GIFs when they wanted their messages to be engaging or they wanted to signal effort in the conversation:

\begin{quote}
\textit{...the person gets that I took some time to find it [the GIF], and then they get, they'll see that, [I spent] more time than just clicking on a smiley face. (P15)}
\end{quote}

While participants used GIFs to express nuanced emotions, make the conversation humorous and playful, and convey engagement and effort in the conversation, we also saw participants describe communication difficulties and breakdowns with GIFs. In the following sections, we discuss factors around GIFs that shaped participants' communication with GIFs, for better and for worse.

\subsection{Technical Functionality}
People's experience of GIF use is shaped by the technical functionality that supports GIFs. Participants' stories revealed that technical overhead associated with GIF use can cause challenges to the presentation of GIFs, which frustrated participants.

\subsubsection{Platform GIF Support}
Participants experienced frustration on platforms that lacked adequate GIF support, including difficulty or even inability to use GIFs. For example, participants were uncertain about using GIFs on some platforms because animated GIFs were not always ``animated,'' as they assumed, which caused miscommunication:

\begin{quote}
\textit{I don't know what the deal is with Facebook. I'm always kinda posting it and it looks like it works, but I close it and it doesn't move, which is an image, then you look stupid...I think there's some particular way just to get the link inserted but I don't know. (P07)}
\end{quote}

\begin{quote}
\textit{If you are doing WhatsApp or something, if you try to paste the GIF it doesn't always animate. Sometimes it's just flat, so it's weird. It's not always as compatible. Whatever is the sentiment or the joke, because that's usually in the motion and if it's just a flat image it's just like that [joke] didn't land. (P12)}
\end{quote}

In P12's case, the fact that GIFs were not animated on WhatsApp directly defeated her purpose of using GIFs to communicate her emotions or jokes, which are expressed over time through the animation. Not only was there no system indication that an animated GIF was the sender's intended presentation, there was also no indication that a static image was not the intended presentation. The probability that the communication partner would correctly understand the presentation therefore drops dramatically.

Even on platforms where GIFs would play properly, inserting them was a complex process that participants found frustrating:

\begin{quote}
\textit{It's kind of a convoluted process...If it were on a desktop, I would... probably find the site that it [the GIF] was on, and possibly get on my phone, go to the browser, and type in the exact URL that the GIF was on from the desktop to my phone, and I can download it to my phone, because you have to save an archived copy of the GIF to post it inline onto GChat [Google Hangouts], or if it was just the link, I would probably cut and paste the URL directly to that image as text into the GChat window. (P08)}
\end{quote}

P08 summarized the support for GIFs on Google Hangouts as \emph{``ridiculously stupid. It is complex. It is absolutely moronic.''} During his interview, P08 mentioned there were times that he still wanted to send a GIF inline, knowing that it required a much more complex process than pasting in the URL of the GIF:

\begin{quote}
\textit{I'm having my inline conversation, I'm talking to this person, I just want to show him this picture, I don't want [them] to click on this thing, I don't want [them] to open up a new app window, I don't want all this. Why can't I do it? It would be immensely more frustrating. But even with that, I think it's still stupid and frustrating to go through that whole process. (P08)}
\end{quote}

According to P08, the overhead of using GIFs can disrupt the conversation flow. The easy way for P08 to insert a GIF is by inserting the link, but that would require the communication partner to open a new window to see the GIF, which would be frustrating. However, even though the process of inserting GIFs inline was more difficult, for P08 it was worth the effort to ensure an engaging conversation.

Using GIFs on platforms without native support is difficult. Participants described elaborate work-arounds to include GIFs, and confusion when GIFs did not behave as expected. One may think that proper GIF support would solve all their problems, but we found that was often not the case. In the next section, we discuss the benefits and challenges participants shared about built-in GIF search engines, a common form of GIF support.

\subsubsection{Built-In GIF Search Engines}
Incorporating a GIF search engine is a common way for platforms to support GIFs. Participants told us platforms with built-in GIF search engines led to increased use.

\begin{quote}
\textit{So I use them occasionally in text messages for a few years, but since I got the new one [iOS, which includes a GIF keyboard]... I've been using it a lot more, because it's just easier to add stuff, instead of having to go to Google and searching something and copying and pasting. (P04)}
\end{quote}

Since iOS 10, iMessage has incorporated a GIF search engine, which makes it convenient to insert GIFs into messages \cite{noauthor_send_nodate}. Similarly, Android's inclusion of a GIF keyboard starting with version 7.1 \cite{noauthor_google_2016} led to P18's increased use:

\begin{quote}
\textit{P18: I have an Android phone, and in my app, my SMS app. I used to just be able to use emojis, but now I can search GIFs...I think it was an update to my SMS app. I used GIFs a lot less in texts before that. }

\textit{Interviewer: Have you been using GIFs more after that update?}

\textit{P18: Absolutely. Definitely.}
\end{quote}

However, even when platforms incorporated GIF search engines, some participants found the search results to be low quality, and they were unable to find the GIFs they wanted:

\begin{quote}
\textit{I don't really look through the app [GIF keyboard] too much because a lot of them aren't of the best quality and they are a little bit short...It's hard to find exactly what I'm looking for. (P03)}
\end{quote}

\begin{quote}
\textit{So I have that GIF keyboard, which made it easy, but I think...its repository of GIFs, I want to say wasn't as compelling...There's a shit load of GIFs, and you go through like you know, I'll go through something like 200 GIFs, before I find like the perfect one...There were just like hundreds of GIFs that were not that good on the keyboard. (P07)}
\end{quote}

Participants' frustration over their inability to find the right GIF again highlighted the importance of conversation flow. Previous online chronemics research found that the chronemic norm of responsiveness favored low response latency \cite{avrahami_responsiveness_2006}, and long pauses between messages could lead to a negative impression and decreased trust \cite{kalman_online_2013,kalman_online_2011}. We see a similar phenomenon with GIFs---one can theoretically take the time to find the perfect GIF, but in high-tempo instant messaging where GIFs are often used, taking the time may not be a feasible option: 

\begin{quote}
\textit{I need to get this message out, but I can't spend forever like, trying to search for the perfect GIF, which I found myself doing a lot of because I'm like, ``Eh, it doesn't really convey what I want to do.'' (P11)}
\end{quote}

\begin{quote}
\textit{When we used to have conversations in GIFs, I would text my friend like, when you sent me that one, I had no clue what that one was for, and she would say like ``oh it's joyful surprise'' but that didn't come through...It's like we've gone past the moment, so the GIF comes later like there's been lag, and so that immediacy, that opportunity for reaction and emotion is missed...when you have to like searching around for it, and say like, someone types something at the meantime, you just sort of miss the opportunity. (P10)}
\end{quote}

Across our interviews, the need to find ``the perfect one'' was expressed by many participants. The definition of ``perfect'' differed from person to person, revealing that the determination is highly context-dependent and shaped by the interpersonal and communal common ground established with the communication partner. In the next section, we discuss in detail the role of context in communication with GIFs, how context shapes the use and understanding of GIFs, and how context can make a GIF ``perfect'' in communication.

\subsection{Context}
In participants' descriptions of how they use GIFs, we found that context played an important role. While technical functionalities often impacted the \emph{presentation} of GIFs, context has a significant impact on the \emph{acceptance} of GIFs. The theme of context was present in every interview. According to the participants, context reflected various kinds of common ground, from personal to communal, and impacted not only what GIFs to use and how GIFs were interpreted by the recipient, but also whether to use GIFs at all.

We identified four different types of context during our analysis: \emph{source material}, \emph{communication partner}, \emph{media norms}, and \emph{platform culture}. Source material refers to the original media from which the GIF visuals originated. Communication partner refers to the interpersonal relationship between the sender and the recipient. Media norms refer to the basic concepts of what a GIF is, how GIFs are used, as well as why GIFs are a legitimate communication form. Finally, platform culture refers to the social practices and culture that can emerge within an online community that uses GIFs. At the end of this section, we discuss the ways in which these types of contexts work together to impact people's use and understanding of GIFs.

\subsubsection{Source Material}
Knowledge of a GIF's source material---that is, the media where the content originated---impacts how people interpret that GIF. A GIF's source material can often give it extra context and consequently additional or completely different meanings than what it means on the surface. We find the context of source material reflects communication partners' communal common ground, particularly shared knowledge of the source material, which typically originates from pop culture such as movies and TV shows.

Knowledge of source material is critical to understanding the sender's intentions in sending a GIF, which can be challenging for people who are not familiar with pop culture. Participants mentioned they did not understand some GIFs because of this lack of familiarity. For example, P07 explained why she did not understand some GIFs that referred to pop culture:

\begin{quote}
\textit{Yeah, usually it's because it's a pop culture reference that I'm not understanding. I think pop culture GIFs are particularly funny, when you're like ``oh it's that character from that show,'' you know, so sometimes you don't have that context, it's kinda like, okay it's a person that's laughing but I don't get it. (P07)}
\end{quote}

P07's confusion over unfamiliar pop culture references highlights how miscommunication can arise when the sender is aware of the reference while the recipient is not, or vice versa. In the first case, the intended extra meaning from the source material is not communicated, and in the second case, an unintended meaning is communicated. Both cases can lead to miscommunication. One participant explained that the second type of potential miscommunication led to their being meticulous about GIF choice in order to avoid the fate of Jane from our opening paragraph:

\begin{quote}
\textit{If I search something, I'm just pulling one down, and it turns out it's a character from that movie, and I send it out to 100 external folks. The movie involved weed humor, so perhaps someone takes offense to that or assumes I'm advocating for that because of using that. I think that's where I am sometimes pretty careful and pretty deliberate with what I use, or more so what I don't use. (P21)}
\end{quote}

Previous work shows that GIFs can be interpreted differently by different people \cite{jiang_understanding_2017}, and here we see a similar phenomenon experienced by participants caused by a lack of communal common ground. P21 avoided using GIFs from movies with marijuana references, because he anticipated that such GIFs would cause miscommunication by conveying unintended meanings from the source material. We also saw further evidence of multiple interpretations from our interviews. As part of our interview protocol, we showed participants GIFs and asked for their interpretations. One showed Homer Simpson saying that he was getting drunk at a flower shop (Fig. \ref{fig:1}); someone familiar with the source material would know that the purpose of this GIF was to illustrate a lie. However, since none of our participants were familiar with that scene from The Simpsons, they interpreted the GIF as an interesting way of saying ``getting drunk,'' and said that they would use it in a humorous, mocking way---but not as an illustration of a lie. If their communication partners had seen the episode, the extra meaning from the source material could be unintentionally carried into the conversations. Similarly, for Fig. \ref{fig:2}, participants who were familiar with JonTron, a YouTube personality, interpreted him as being excited, while those who were unfamiliar interpreted him as being angry.

Our findings show that miscommunication can happen if there is a discrepancy in source material knowledge among communicators. Across our interviews, we also saw participants selectively send or not send GIFs based on their knowledge about their recipients, which we discuss in the next section.

\subsubsection{Communication Partners}
The context of the interpersonal relationship between communication partners shapes how people use GIFs. Participants actively leveraged their personal common ground and assessed whether and how GIFs would be understood by the recipient, and miscommunication under this type of context also tended to happen when their communication partners did not understand the meanings of GIFs. A simple rule of thumb that participants followed was to send GIFs that they were confident their communication partners would understand, and not to send those about which they were less confident. For example, P04 told us how she selected GIFs to send to one of her friends based on that friend's preferences and presumed knowledge:

\begin{quote}
\textit{[L]ike my one friend, she likes animals so I send her animal GIFs, but she doesn't use the Internet as much I think, so I wouldn't send her GIFs with a meme, probably. I wouldn't know if she knew the meme or not. It'll be kind of random to send her like, some GIF and she'll be like ``what is that,'' and I'll be like ``oh, never mind.'' (P04)}
\end{quote}

P04 would not send meme GIFs to this friend because she did not feel confident that her friend would understand the memes, and anticipated miscommunication if she sent such GIFs. Instead, she would send the friend animal GIFs to cater to her preferences and avoid questions. This example, together with P21's example in the last section, reveals a way that personal and communal common ground often work together to shape the sender's choice of GIF. In both cases, the sender was unsure about whether the recipient would understand the cultural reference, which results from a lack of knowledge about their communication partner, and eventually took the safe route to avoid miscommunication.

Participants also took advantage of their knowledge about their communication partners, and used GIFs for which communication partners have a shared context to enhance the communication. P09, for example, explained how this shared context could bring communication to ``another level'':

\begin{quote}
\textit{I sent her some GIFs from Lion King, and it was like the baby Simba...and he says ``You're so weird...'' It's one thing I can just tell somebody, like if I just say you are weird, but we both have seen the movie Lion King, and both know the context of the movie, then it adds like, another level to the communication. (P09)}
\end{quote}

Besides referring to shared knowledge of source material, GIFs can also refer to shared real-life experiences of communication partners, which participants referred to as ``inside jokes.''

\begin{quote}
\textit{Sometimes my neighbor has really loud sex, so I just sent her ``congrats on the sex'' GIF, with the cake, with nothing else, and just wait for her to reply with like ``oh my god I'm so sorry. It'll never happen again...'' I'll just send her that GIF, and like, she knows what I'm talking about, and she knows why I'm annoyed. (P17)}
\end{quote}

Here, the shared experience between communication partners helped them add meanings to the GIF. We were surprised to find that the power of the context of communication partners extended beyond enhancing communication---sometimes it could also overcome technological difficulties mentioned in the last section. Some participants chose to simply use words to describe the GIFs that they wanted to send when they had trouble finding them in search engines.

\begin{quote}
\textit{Sometimes because we've used these [GIFs] so often, that we'll just like, put in brackets, like the title of the GIF that we use, like the description of the GIF that we would've used in that position, that point, and the other person just totally knows what it means. (P10)}
\end{quote}

\begin{quote}
\textit{If it's the common ones that we both use...sometimes you don't need to put the actual GIF if I can just say like ``such and such GIF'' and she knows what it is. (P19)
}\end{quote}

While most participants thought that GIFs were more expressive than words in their description of the appeal of GIFs, here we see some participants used words to \emph{replace} GIFs. While people may prefer GIFs over other nonverbal communication media because of the level of nuance they can convey, people can use text as short-hand to reference familiar GIFs when they communicate with the same GIFs frequently, rather than spending the time to find and insert them. We find this practice consistent with Clark and Schaefer's theory of least collaborative effort, in which communication partners try to minimize the total effort spent in conversations \cite{clark_contributing_1989}. Moreover, this practice echoes our previous discussion on the difficulty and overhead associated with using GIFs---if a shorthand sufficiently communicates the context, there is no need to spend the effort on finding the right GIF.

The context of communication partner shapes not only the choice of GIFs, but also the choice to use them at all, which we saw more commonly in anticipated communication between participants and their parents. For example, P03 explained to us why he would not send his parents GIFs:

\begin{quote}
\textit{Just because I don't think they would get it...It's kind of a niche online humor that if you are not part of it, you are not gonna get it, so there's no point in sending it to them. (P03)}
\end{quote}

Like P03, many participants mentioned the tension between their habit of using GIFs and their parents' lack of understanding. However, participants believed their parents' confusion was not with the content portrayed in GIFs, but with the concept of GIFs themselves. The concept of GIFs, along with conventions of use, is the third type of context that overwhelmingly led to non-use of GIFs, which we discuss in the next section.

\subsubsection{Media Norms}
While the last two types of contexts focus on the content of GIFs and determine what kinds of GIFs to use, we found another type of context that shaped the decision whether to use GIFs at all. This type of context, which we define as media norms, reflects a basic understanding of the concept of GIFs, which is invisible in experienced GIF users' communication, but arises in communication with people who are not as knowledgeable about GIFs. Media norms consist of three concepts: what GIFs are, how they are used, and why they are a legitimate form of communication. Participants described communication breakdown related to media norms, but it differed from other types of context: In this case, the recipient received a piece of media, but did not understand what it was or why it was used in communication.

Knowledge of technology purpose and practice can commonly differ on dimensions such as age, culture, and education \cite{rogers_diffusion_2003}. We found that the knowledge gaps here are perceived by participants as largely generational---twelve participants had never sent their parents any GIFs due to believing that their parents lacked an understanding of common practices around GIFs, or even a basic understanding of what a GIF is.

\begin{quote}
\textit{I just don't think they would get the use of GIFs...You throw in a GIF, and that's just an occasion they wouldn't understand what it's for, what it is. (P19) }
\end{quote}

\begin{quote}
\textit{I think GIFs will just confuse her more than anything. I don't send GIFs to her at all...She would probably have questions like ``What is that?''  ``Did you mean to send that to me?'' (P20)}
\end{quote}

Another problem participants anticipated when considering sending their parents GIFs was that their parents would not understand the convention of using GIFs to communicate at all---``Why are GIFs a legitimate form of communication?''

\begin{quote}
\textit{[L]ike our parents don't really...It's just they don't understand quite as well, or grasp the concept of GIFs...The form of the GIF and why. Why did you send me that? I think that's what they don't understand too. (P05)}
\end{quote}

\begin{quote}
\textit{[P]arents aren't necessarily as familiar with Internet communication norms. The idea that you can have a conversation using GIFs is something my parents are not comfortable and familiar with. (P18)}
\end{quote}

In some more extreme cases, participants thought their parents would hold negative views about GIFs, and thought GIFs were not just an unfamiliar, but wrong way to communicate.

\begin{quote}
\textit{My mom. My dad. I wouldn't send them GIFs...They would view it as the continued destruction of the English language or something. Like they would see it as not a proper way to communicate... [My mom] would probably tell me to stop. She sees it as an unintelligent way of communicating. (P22)}
\end{quote}

Finally, the common way people use GIFs to convey emotions is also a concept that participants thought would confuse parents. People often use GIFs excerpted from TV show or movie clips that resemble the senders' actions or emotions as a form of virtual self-representation. Tiidenberg and Whelan characterized this kind of use as GPOY, or ``gratuitous picture of yourself.'' \cite{tiidenberg_sick_2017} However, some participants believed their parents would have trouble making the connection between the visual and the sender.

\begin{quote}
\textit{They [my parents] don't really understand GIFs...they would get it just like, oh the person's doing something, but I feel like they wouldn't completely get that it's supposed to be like me doing it. I just don't think they would make that connection...like why is this person doing this, what does it have to do with what we are talking about. They wouldn't have made that connection that it's kind of like a reaction GIF. (P19)}
\end{quote}

While participants anticipated confusion and questions their parents would have, they did not express the intention to explain the common practices around GIFs to their parents. For P17, the choice to not send GIFs to her parents was the result of her trying to avoid questions:

\begin{quote}
\textit{[I]f I anticipate getting a lot of questions like ``Who is this man?'' ``Why is he doing that?'' ``Why did you send this?'' ``What is it supposed to mean?'' then I won't send it...I wanted to further the conversation, not cause them to derail it. (P17)}
\end{quote}

This avoidance of answering questions suggests a potential obstacle to using a novel communication media. While it may seem desirable to use a new communication technology, it also comes with the obligation to explain the technology to the communication partner, particularly when there is a lack of shared knowledge of the technology due to generational or cultural differences. The reluctance to fulfill this obligation may lead to non-use of the communication technology.

While there are some media norms among GIFs, different platforms also have their own practices that are reflected in their norms and cultures. Platform culture as a type of context also shapes how people use GIFs, as we describe in the next section.

\subsubsection{Platform Culture}
Perception of a platform culture can influence the choice of whether to use GIFs on that platform. Specifically, participants considered social practices and culture that they observed on the platform and compared them to their own communication practices. For example, participants preferred to use GIFs on platforms and in communities where people used GIFs as the primary media of communication: 

\begin{quote}
\textit{I think that [GIFs are] part of like, the culture of Tumblr itself. The use of GIF to communicate. (P09)}
\end{quote}

\begin{quote}
\textit{Reddit is the home of GIFs...There are entire subreddits devoted to GIFs, like ones on all cat GIFs, or ones on making GIFs out of Buzzfeed...It's a style of content creation. (P18)}
\end{quote}

Besides common media of communication being important, participants also liked communities with styles of communication where GIFs fit well:

\begin{quote}
\textit{I feel like people are having the kinds of fast-paced conversations [on Twitter] that GIFs lend themselves to. (P12)}
\end{quote}

P12 used GIFs the most on Twitter because she thought they were good for the fast-paced conversations common on that platform. However, some participants chose not to use GIFs on Twitter also because of communication style, like P13:

\begin{quote}
\textit{I take precision of language pretty seriously...the same principle applies to my use of GIFs. It takes me a while to find the right one, and that's not always consistent with Twitter's ethos, I guess. It's a little more ephemeral than that. (P13)}
\end{quote}

This is an example of incongruence between personal communication practices and the dominant communication style of a platform. Many participants preferred platforms that better matched their own communication styles. For example, P08 preferred to use GIFs on Facebook over other platforms because it had a larger audience:

\begin{quote}
\textit{I probably use Facebook or Facebook-related products for GIFs than GChat or Twitter DMs...It's because of the larger audience...Because there's a larger audience, there's just more conversational movement...So more people, more movement; more movement, more use for GIFs there is...I would probably intend that [GIFs] more for a broader consumption, because if it's just someone you're talking to personally, I think I can hit a better level of specificity by just typing. (P08)}
\end{quote}

On the other hand, P10 preferred to use GIFs for direct communications and elected to not use GIFs on platforms where communication partners are more separated:

\begin{quote}
\textit{I would say I use them [GIFs] probably on Hangouts the most...I feel like Hangouts is really meant for communication. It is really meant for me to have conversation and direct communication with people on the other line. I don't actually think that Tumblr is meant for that...Hangouts is more immediate. Tumblr to me is...more removed, more curated, more separated from the person I am talking to, or I am engaging with. (P10)}
\end{quote}

While P09 considered GIFs as part of the culture of Tumblr, P10 preferred Google Hangouts, where it was difficult to use GIFs as we mentioned in the Platform GIF Support section. While a GIF-supportive culture is important, choosing a platform is also highly personal. Congruence with personal communication style plays an important role in choosing platforms to use GIFs.

In sum, our findings show that context significantly shaped participants' use and understanding of GIFs. Understanding of the content of GIFs depended on the context of source material and communication partner. Understanding of the fact that GIFs can be used to communicate, however, required the context of common practices, which participants saw as generational. While participants considered whether and how their communication partners would understand the GIF on an individual level, they also took into account community cultures and how well a platform matched their own communication style.

\section{Discussion}
Through the stories we heard, we found that participants had complicated communication practices and experiences with GIFs. While participants try to find the ``perfect one'' to portray a wide range of nuanced emotions, our data highlights the challenges participants encountered as well, both in sending and in interpreting GIFs. Viewed through the lens of Clark's common ground, these challenges can be seen as the result of inadequate shared understanding. People want to find the perfect GIF that reflects established common ground with their communication partners, but miscommunication with GIFs often happens as a result of inadequate context of source material, interpersonal relationship, or the media itself. In this section, we first discuss the benefits and challenges of using GIFs, and then discuss design implications for nonverbal communication where shared context is salient. Building on our analysis, we argue that the design space for GIFs should not be limited to system interactions and mechanics alone. Rather, interpersonal contexts are so central to use and interpretation that they should be a primary consideration when studying and designing interactions with GIFs.  

\subsection{The Benefits and Challenges of GIFs}
While participants used GIFs in similar ways to emoji \cite{cramer_sender-intended_2016,kelly_characterising_2015}, we also found that participants appreciated how GIFs could convey more complex ideas and emotions, and how they signaled effort and engagement in the conversation. Participants tried to find the perfect GIFs to communicate not only these GIFs' evident meaning on the surface, but also personal common ground, such as ``inside jokes'' or unique experiences specific to communication partners, and communal common ground such as a reference to a movie that they watched together or belonging to the same community. The shared contexts in GIFs became especially salient in times of technology failure. When unable to find the perfect GIF in search engines, participants still managed to refer to the GIF they wanted by using text as a short-hand to convey the shared context, leveraging the personal common ground of a frequently used GIF with that communication partner.  The ability of GIFs to communicate shared contexts may also explain why they can take communication to ``another level.'' 

On the other hand, the multiple contexts we have discussed that GIFs leverage also limit who can participate in and understand communication with GIFs. While previous work showed miscommunication can result from technical incompatibility, for example, interpretations of emoji might vary because the same emoji can look different on different platforms \cite{miller_blissfully_2016}, we found that miscommunication of GIFs often resulted from misinterpretation of contexts and inadequate common ground. For example, participants anticipated unsuccessful GIF communication with older generations due to the lack of communal common ground of what a GIF is and why GIFs are used in communication. Without knowing the concept of and norms around GIFs, older generations would notice that something was said. However, they would not understand GIFs as a popular way to communicate and that what a GIF portrays is ``supposed to be like me doing it.'' This lack of understanding would therefore cause potential communication breakdown. 

Even those familiar with GIFs and who know how it is used can experience miscommunication with GIFs due to a lack of context of source material and interpersonal relationship. Jane's story demonstrates this kind of miscommunication with GIFs: while Jane thought she was expressing happiness, her friend interpreted the GIF with the deeper context of source material that references marijuana use. In this case, the misinterpretation of context is caused by both a lack of communal common ground---Jane did not know the GIF's cultural reference, while the communication partner did---and a lack of personal common ground---the communication partner did not know Jane's unfamiliarity with the cultural reference. 

Similarly, participants shared their confusion about cultural references in GIFs. Successful communication with GIFs often required communication partners to have established personal and communal common grounds in order to perfectly align multiple layers of contexts, which can be difficult, as participants' stories showed. At the same time, because GIFs are interpreted through multiple contexts, they can be interpreted in more diverse ways and are easier to misinterpret under a lack of common ground. This challenge of GIFs can easily confuse people unfamiliar with GIFs. Our interviews focused on those familiar with GIFs, and as a result we can only speculate on how non-users of GIFs feel when receiving GIFs. However, we do know that our participants limited their use of GIFs to communication partners they thought would understand GIFs, effectively limiting who can benefit.

As a context-dependent media, GIFs are a double-edged sword. Sending GIFs with shared contexts, which comes from established common ground, can enhance communication, while using GIFs without shared context can be detrimental. A potential way to deal with this tradeoff is for people to only use a GIF when they are certain their communication partners have a common ground through which to interpret the image. However, such meticulousness may not be compatible with either the fast-paced conversations that GIFs lend themselves to or the purpose they are typically used for---to make conversations interesting and funny. These affordances and challenges suggest GIFs are a uniquely nuanced form of nonverbal communication.

\subsection{Communicating Context}
Compared to other communication media such as text or emoji, GIFs derive much of their meaning from context. While emoji's meanings can also be impacted by context, emoji as a set of generic symbols do not make their context salient; GIFs, on the other hand, actively invite people to go beyond their obvious meanings and discover deeper meanings that come from multiple layers of context. Beyond the technical and cultural savvy on which GIFs rely, participants discussed how specific interpersonal relationships shaped their use (and non-use) of GIFs, and ultimately much of their affective appeal.

Miltner and Highfield \cite{miltner_never_2017} argue that decontextualization of GIFs from their original source material results in a polysemic quality, enabling a variety of different interpretations for different communities. Our analysis supports their claims, but also highlights the importance of a subsequent step: re-contextualizing GIFs within specific interpersonal relationships. 
Interpersonal context is not lost in GIFs or displaced by the source material. In talking with people about their experiences and practices, we find it is through the sharing of GIFs that interpersonal contexts can be established, deepened, and reinforced. The polysemic GIF, when used within an interpersonal context, can suture the conversation partners around an established shared understanding of the GIF that can be specific to that interpersonal relationship. Just as memes gain additional layers of meaning as they circulate, GIFs can gain meaning distinct to the specific interpersonal context. Just as memes create in-group/out-group distinctions for those ``in the know,'' GIFs provide a flexible and highly appropriable channel for communication partners. 

Sociolinguistic scholarship suggests language does not simply communicate information but also establishes and maintains relationships \cite{trudgill_sociolinguistics:_2006}; we found people used GIFs in the same way. In many cases, participants described using GIFs to communicate precisely even though the media often provides ambiguity. While previous work has focused on the affordances and richness of the media in expression and interpersonal communication \cite{bakhshi_fast_2016}, we found participants both leveraging and reproducing the interpersonal context through GIFs. 

The affective appeal described by our participants, then, is based on more than just the expression of emotions. It is also the result of familiarity and intimacy that GIFs perform within the interpersonal relationship, and through which they gain their significance. For these reasons, interpersonal context should be a primary consideration when designing for GIFs.

\subsection{Designing for Context}
The challenges participants faced in using GIFs suggest design opportunities for nonverbal communication where shared context is salient. However, our findings suggest that the design solution might be more complex than simply adding a feature. Participants were frustrated by the complex processes required to insert GIFs on platforms without GIF search engines, and the underwhelming search results on platforms that have them. An intuitive solution would be to incorporate GIF search engines into those platforms, and to rank commonly used GIFs higher for potentially better search results. However, participants wanted the GIF to be ``perfect''---perfect for the communication partner, the emotion, and the platform, and our findings suggest that it is the shared context that makes a GIF ``perfect.'' In other words, different GIFs are appropriate for different conversations and interpersonal relationships. Communication partners and community cultures can also imbue GIFs with meanings that differ from their common interpretations. These nuances in communication may apply to areas other than GIFs and be generalized to the broader genre of nonverbal communication. 

The importance of context reframes the design space of this media type to include consideration of the context itself. Therefore, while supporting transactional needs when communicating with GIFs (e.g., search keyboards) would certainly be well received, the full design space (albeit a more challenging space)  should include the complex social contexts and common grounds that people have, besides technological features that would be easy to use. We can consider design opportunities that support different kinds of context and common ground: to support contexts of communication partners, GIF keyboards may want to recommend GIFs frequently used with each other to leverage the personal common ground, rather than across all conversations. To support context from source material or platform culture, designs could crowdsource and show different explanations of GIFs in different communities where the GIFs are used, leveraging communal common ground.  

To support the context of media norms, we want to point out that while we would expect that this type of context would be the easiest and the most intuitive for people to adopt, design solutions beyond making the GIFs available and encouraging people to use them are difficult. While a simple example, a design opportunity could be a tooltip that prompts people to send a GIF when they receive one, and then prompts them GIFs under emotion categories to show that emotions can be associated with a media type like GIF.

\subsection{Limitations \& Future Work}
There are some limitations to this study. First, we want to emphasize that many of our statements about participants' communication partners, especially their parents, are based on the participants' perceptions. We also only conducted interviews with experienced GIF users, who may have a biased view in favor of GIFs. A closer look at the experiences and practices of communication pairs, parents, as well as inexperienced GIF users will provide a fuller picture of the nuances of communication with GIFs, which we leave for future work. 

Second, this work does not address the impact of cultural contexts, which prior work has shown to impact nonverbal communication media other than GIFs \cite{lu_learning_2016,park_cross-cultural_2014,park_emoticon_2013}. Therefore, our future work also includes an investigation of how cultural contexts impact communication with GIFs.

One of our goals in our initial, exploratory work was to uncover the interesting insights that could be further explored in future CSCW work that could go deeper on individual aspects of GIF use, such as under certain interpersonal context or platform context---how do romantic partners use GIFs? Or how has GIF use on Reddit changed over time? We hope that our findings help motivate additional contributions in this space.

\section{Conclusion}
In this exploratory study, we examined the nuances and challenges in people's communication experiences with GIFs through in-depth qualitative interviews. We found that participants' communication with GIFs was impacted by technology and context around GIFs---both for good and for bad.

This research provides two main contributions. The first contribution lies in our deeper understanding of people's miscommunication experiences with GIFs, situated with their nuanced communication practices and experiences with the novel communication media. While people used GIFs for their strong expressive power, the contexts that GIFs convey also provide paths to miscommunication when there is a lack of personal or communal common ground. We also contribute design implications for nonverbal communication media that rely on shared contexts, suggesting that accounting for people's social context and common ground is just as important as designing for easy access to a communication feature.

\begin{acks}
We would like to thank the reviewers for their significant time and care in improving this paper. We would also like to thank Jeremy Birnholtz, Colleen Jankovic, Bryan Semaan, and the HCC community at CU Boulder for their immensely valuable feedback on the paper. The first author thanks Arcadia Zhang for her endless support and feedback during the paper writing process. Finally, we would like to thank all GIF users and especially our participants, who have made GIFs such a fascinating topic to study.
\end{acks}

% Bibliography
\bibliographystyle{ACM-Reference-Format}
\bibliography{gif}

\end{document}